\documentclass[graybox]{svmult}
\usepackage{graphicx} 
\usepackage{amsmath, amsfonts, amsthm, amssymb}
\usepackage{lmodern}
\usepackage[T1]{fontenc}
\usepackage{mathrsfs}
\usepackage{xcolor}
\usepackage[numbers]{natbib}
\usepackage{hyperref, caption}

\def\vnk{$v_{\rm NK}$}

\newcommand{\rev}[1]{{#1}}

\begin{document}

\title*{\rev{Compact Object} Astrophysics with Frontline Astrometry}
\author{Poshak Gandhi}
\authorrunning{P. Gandhi}
\institute{School of Physics \& Astronomy, University of Southampton, Southampton SO17~1BJ, UK.\\
Inter-University Centre for Astronomy \& Astrophysics, IUCAA, Post Bag 4 Ganeshkhind, Savitribai Phule Pune University Campus Pune 411 007, India.\\
\email{poshak.gandhi@soton.ac.uk}}
\maketitle

\abstract{Astrometry --- the precise measurement of celestial positions and motions --- is entering the micro-arcsecond ($\mu$as) era at multiple wavelengths, enabling new insights on compact objects across all mass scales. Here we review how high-precision astrometry is advancing our understanding of \rev{compact objects} --- neutron stars (NSs) and black holes (BHs). We provide the context for high precision astrometry before discussing natal kicks and the latest results from \textit{Gaia} Data Release 3 (DR3). We highlight the evidence for mass-dependent peculiar velocities of accreting binaries, and also reveal a close similarity between NSs and BHs. Next-generation surveys will find recoiling supermassive BHs (SMBHs) in galactic nuclei, exploring how gravitational-wave-induced kicks operate. Exploitation of scientific opportunities on the lunar surface could facilitate much larger collecting areas and astrometric precision in X-rays than currently feasible.
}

\section{High-Precision Astrometry \rev{Parameter Space}}
\begin{figure}
\centering
\includegraphics[width=0.95\textwidth]{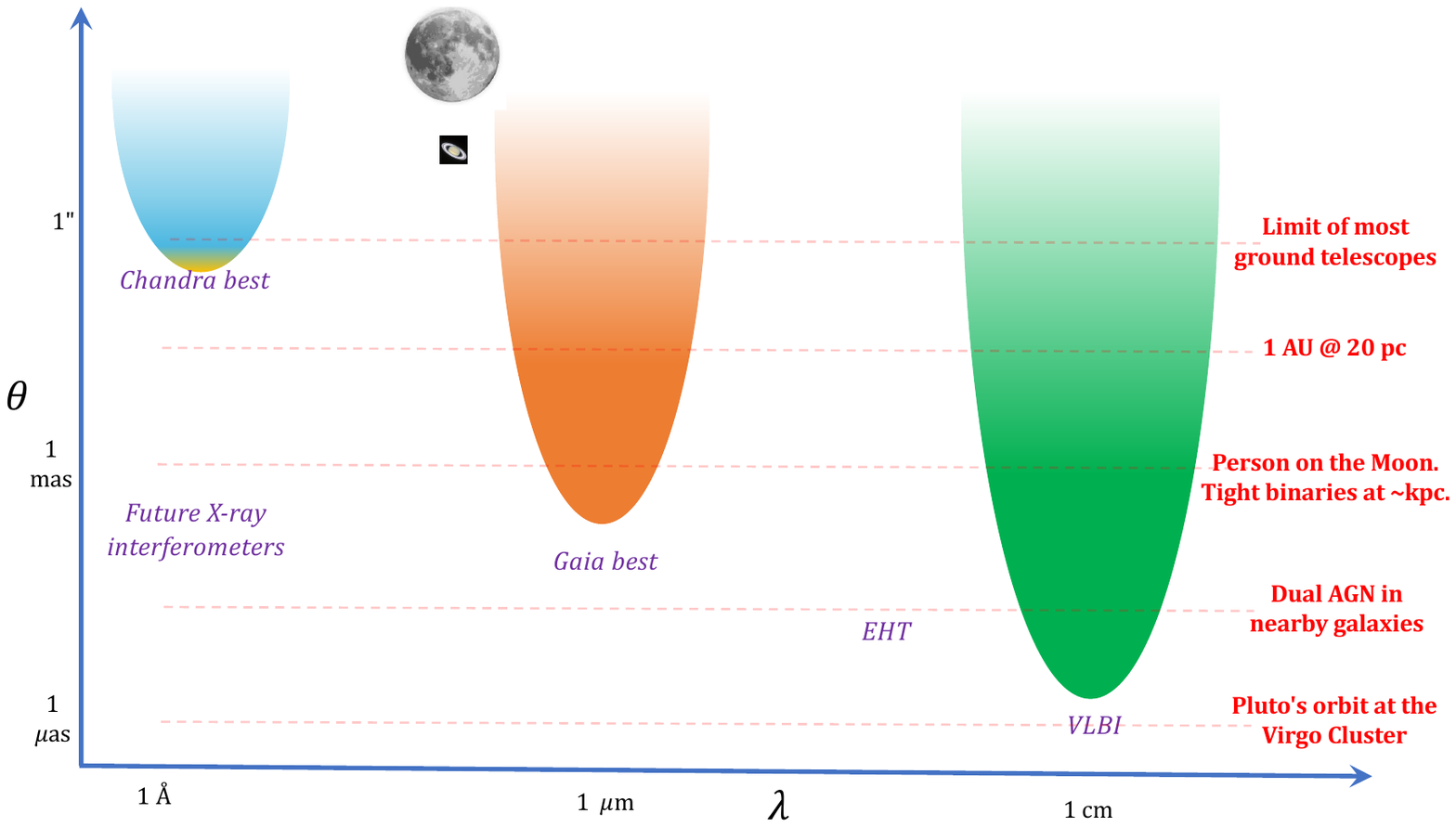}
\caption{Astrometric precision scales spanning several orders of magnitude in angle. \rev{The figure shows characteristic angular scales as a function of wavelength, with some familiar corresponding sizes annotated on the right.} Early optical astronomy was limited to $\sim$arcsecond accuracy. Modern techniques achieve milliarcsecond (mas) precision (e.g. \textit{Hipparcos} at $\sim$1~mas) and even microarcsecond ($\mu$as) precision (e.g. \textit{Gaia} and VLBI astrometry). This improvement enables new science, such as detecting tiny stellar wobbles and high-velocity compact objects. The approximate angular scales spanned by the full Moon and Saturn in the optical are shown for comparison.}
\label{fig:astrometry_scale}
\end{figure}

Astrometry is among the oldest branches of astronomy, but only recently have we reached the precision required to address cutting-edge problems in compact-object physics. Traditionally, ground-based optical astrometry was limited to $\sim$1\,arcsecond resolution by atmospheric seeing. The advent of space observatories (e.g. the \textit{Hipparcos} satellite) improved this to the milliarcsecond (mas, $10^{-3}$~arcsec) level, and today the \textit{Gaia} mission has achieved parallax and proper motion accuracies down to tens of micro-arcseconds ($\mu$as) for millions of stars \cite{gaiamission}. For context, $1~\mu$as is an exceedingly small angle, roughly the apparent size of a mosquito as seen at the distance of the Moon. Ground-based very long baseline radio interferometry (VLBI) can likewise attain $\sim 10$~$\mu$as relative astrometric precision on bright radio sources \cite{reid14_review}, rivalling \textit{Gaia} on select targets. Fig.\,\ref{fig:astrometry_scale} shows the current parameter space of approximate astrometric precision as a function of observing wavelength. These advances mark a transformative jump in our ability to chart celestial motions.

At sub-mas precision, one can begin to measure the annual parallaxes of stars across much of the Milky Way, mapping their distances and three-dimensional space motions. Pushing into the $\mu$as regime opens up other applications: detecting the minute wobble of stars induced by orbiting planets or unseen companions \cite{gaiabh3}, measuring the tiny deflection of starlight by compact masses \cite{sahu22}, and accurately tracking objects moving at high velocities \cite{brown15}. In the context of compact objects --- neutron stars (NSs) and black holes (BHs) --- $\mu$as astrometry enables us to measure proper motions (transverse velocities) of these exotic remnants even at large distances \cite{zhao23}. When combined with line-of-sight radial velocities, the true space velocities of these objects can be reconstructed, offering clues to their origins and dynamics in the Galaxy. 

\section{Natal Kicks and Compact-Object Kinematics}
Massive stars are typically born as part of binary systems, and end their lives as compact remnants --- NSs or BHs --- often via core-collapse supernovae. These violent birth events can impart a recoil momentum to the newly formed compact object, a phenomenon known as a \emph{natal kick}. 

\rev{The binary will remain bound if the ejected mass ($\Delta M$) in the supernova (SN) is less than half of the total progenitor system mass before SN. The systemic velocity recoil imparted to the system's centre-of-mass will manifest as a `peculiar' motion relative to typical Galactic stellar motions, and can be expressed using Kepler's Third Law as}

\begin{equation}
\footnotesize{
v_{\rm pec} = 213\ \mathrm{km\,s^{-1}} \times 
\left( \frac{\Delta M}{M_\odot} \right)
\left( \frac{M_2}{M_\odot} \right)
\left( \frac{P_{\rm postSN}}{\mathrm{day}} \right)^{-1/3}
\left( \frac{M_{\rm tot}}{M_\odot} \right)^{-5/3} 
(1 - e_{\rm postSN}^2)^{-1/2}}
\end{equation}

\noindent
\rev{with $P_{\rm postSN}$, $e_{\rm postSN}$, and $M_{\rm tot}$ being the orbital period, eccentricity and total system mass, respectively (all post-SN), and $M_2$ the companion mass  \citep{nelemans99}. These recoil velocities have come to be known in the literature as `mass-loss' kicks, or symmetric `Blaauw kicks' \cite{blaauw61}. They set a useful baseline to expected peculiar motions of kicked sources.} When a binary system with a compact object and an ordinary companion star survives a supernova, astrometry provides a direct window into these kicks by tracking the proper motions and distances of the companion, thereby yielding their space velocities.

Observational surveys of pulsar proper motions have long indicated that NSs receive substantial natal kicks. \rev{Hobbs et al. \cite{hobbs05}}, for example, analyzed $\sim 200$ radio pulsar velocities and found a broad distribution well fit by a Maxwellian with one-dimensional dispersion $\sigma \approx$\, 265\,km\,s$^{-1}$ (and see Disberg \& Mandel  \cite{disberg25} for an update on this). This implies a significant fraction of NSs are born with \rev{natal-kick speeds (\vnk)} of hundreds of km~s$^{-1}$, and some with $\sim$\,1,000\,km\,s$^{-1}$ or more, \rev{sufficient to perturb them onto highly eccentric, or even unbound, Galactic orbits}. Such strong natal kicks are generally attributed to asymmetries in the supernova explosion (whether in ejecta or neutrino emission) \rev{over and above Blaauw kicks \cite{janka13, popov25}}. High-precision astrometry has been essential in firming up these results; e.g., measuring a 10\,mas\,yr$^{-1}$ proper motion for a pulsar at a distance of 2 kpc (say) translates to a transverse speed of $\sim$\,95\,km\,s$^{-1}$. Today, dozens of pulsar proper motions are known to better than 1 mas\,yr$^{-1}$ accuracy, refining the statistical kick models \cite{bray18, taurisbook}.

The situation for black holes is more complex. Stellar-mass BHs can form via similar supernova mechanisms (in the case of lower-mass progenitors that still undergo explosions) or via direct collapse of a massive star with little explosive ejecta. In the latter scenario, one might expect a negligible kick, since there is little mass expelled to provide any impetus to the BH. Indeed, some observational evidence supports the existence of BHs with very small natal kicks. A notable example is the BH in the X-ray binary VFTS\,243 in the Large Magellanic Cloud, which appears to have formed via direct collapse with essentially no kick --- the BH remains in a circular orbit around its O-star companion within a young stellar cluster, inconsistent with a large natal recoil \cite{shenar22}. \rev{Similarly, the prototypical Galactic low-mass BH X-ray binary V404\,Cygni appears to reside in a hierarchical triple system strongly limiting the maximum kick that it could have survived \cite{burdge24}.}

On the other hand, there are BH systems that clearly did experience substantial natal kicks. X-ray binaries hosting BHs in low-mass stellar systems (LMXBs) often show large space velocities or high Galactic latitudes. A well-known case is the BH LMXB XTE~J1118+480, which has a high proper motion and lies far ($\sim 1.6$~kpc) above the Galactic plane, implying a birth kick of order $\sim 200$~km~s$^{-1}$ if it originated in the disk \cite[e.g.][]{mirabel01}. \rev{In more systematic studies, Repetto et al. \cite{repetto12} found that several known BH LMXBs are observed at heights well above the thin disk, and they argued,} that this is incompatible with formation in situ – natal kicks are required to explain their present orbits. Such results suggest a dichotomy: some BHs (perhaps those formed from the most massive stars) form via direct collapse with minimal kicks, whereas others (from somewhat lower-mass progenitors or asymmetric fall-back events) do receive kicks, albeit possibly on average smaller than those of NSs. This is further supported by a study showing that tight (short-period) binaries are preferentially scattered substantially above and below the Galactic plane, as opposed to longer-period systems that are concentrated closer to the plane \cite{gandhi20}; it is the tight binaries that are able to survive strong kicks and withstand disruption. 

Most such studies have historically been hampered by small number statistics. This is now changing thanks to the latest astrometric data from \textit{Gaia}.

\section{\textit{Gaia} DR3 Findings on Compact Objects}
The third data release (DR3) of the \textit{Gaia} mission has provided an unprecedented trove of astrometric data, including parallaxes and proper motions for over a billion stars, and radial velocities for many millions \cite{gaiadr3}. \textit{Gaia} has measured the proper motions of numerous X-ray binaries (via their optical companions) and, more recently, non-interacting compact-object systems, enabling a systematic kinematic census; see also \cite{gandhi19, atri19, arnason21}. In recent work, Zhao et al. \cite{zhao23} compiled a catalogue of 85 compact-object binaries (encompassing both NS and BH candidates in high- and low-mass X-ray binaries, as well as radio pulsars in binary systems) with available \textit{Gaia} DR3 proper motions and (when possible) systemic radial velocities. By integrating their orbits in a Galactic potential, they directly inferred the 3D peculiar velocities $v_{\mathrm{pec}}$ for each system which, in turn, is a proxy for the natal kick velocity \vnk.

Zhao et al. find that the overall $v_{\mathrm{pec}}$ distribution of the 85 systems is bimodal. It can be described by a two-component Maxwellian, with a low-velocity component having dispersion $\sigma_v$ \,$\approx$\,21\,km\,s$^{-1}$ and a high-velocity component with $\sigma_v$\,$\approx$\,107\,km\,s$^{-1}$. In other words, there is a substantial subpopulation of compact binaries with very small peculiar velocities (tens of km\,s$^{-1}$, comparable to that of ordinary stellar populations), and another with much larger velocities (hundreds of km~s$^{-1}$) consistent with classic natal kick scenarios. When Zhao et al. divide the sample into categories, an interesting pattern emerges: binaries with high-mass, typically early-type donor stars (e.g. high-mass X-ray binaries) mostly occupy the low-velocity group, with $v_{\mathrm{pec}}$\,$\lesssim$\,100\,km\,s$^{-1}$, cf. Nuchvanichakul et al. \cite{nuchvanichakul25}; whereas those with low-mass companions, cf. pulsar binaries or BH LMXBs \cite{dashwoodbrown24} span a broad velocity range extending up to $\sim$\,400\,km\,s$^{-1}$. This suggests that systems involving more massive stars in a binary tend to receive smaller kicks, consistent with momentum recoil expectations. Lighter systems, many of which host NSs, show evidence of the full kick spectrum up to very large recoil speeds.

\begin{figure}
\centering
\includegraphics[width=0.75\textwidth]{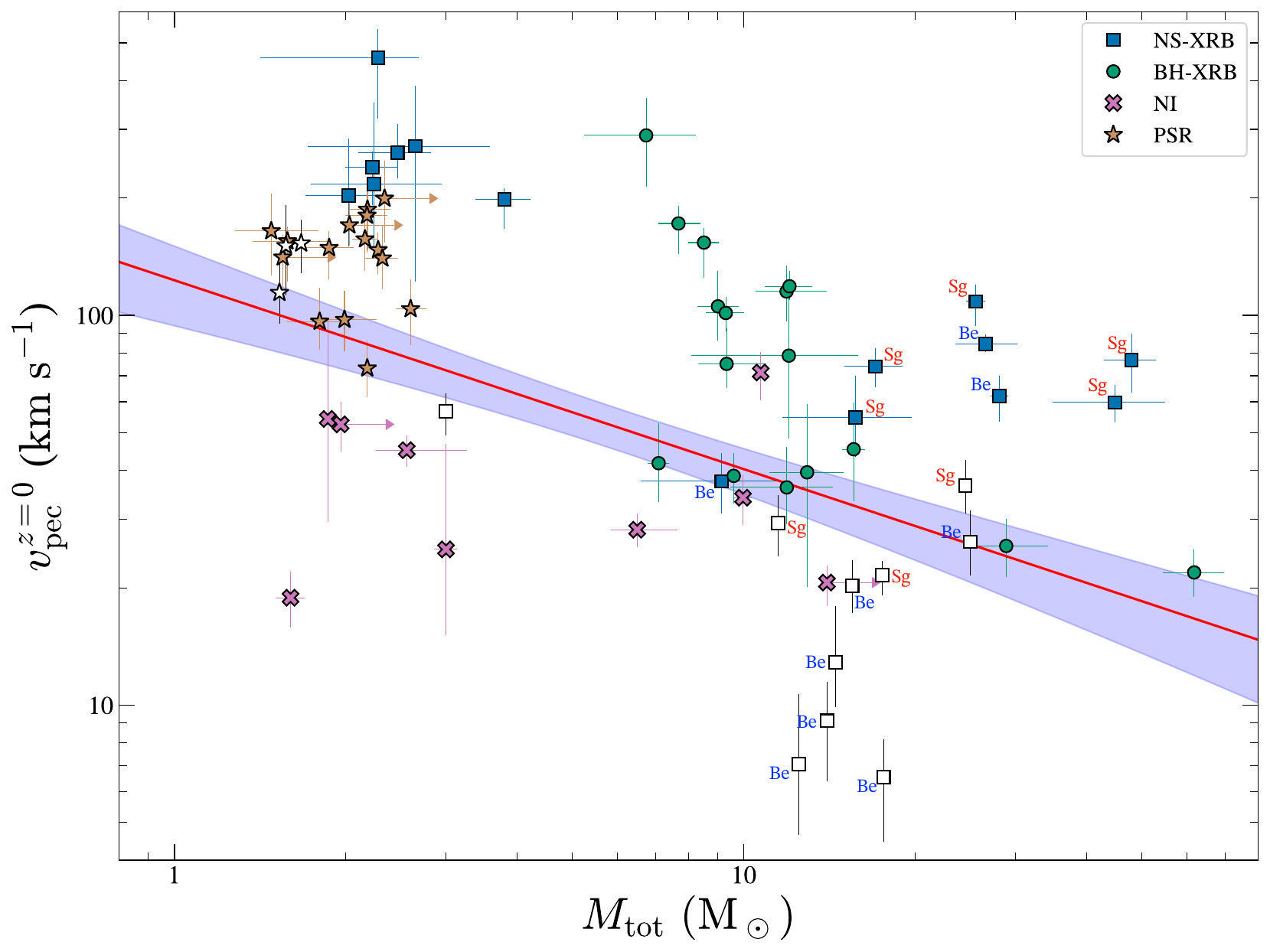}
\caption{Mass-dependent peculiar velocities of compact-object binaries from Zhao et al.\rev{ (Reproduction from Ref. \cite{zhao23}).} The trend shows that binaries with higher total system mass (often those hosting black holes with massive stellar companions, towards the right) tend to have smaller peculiar velocities, whereas lower-mass systems (e.g. neutron star binaries with low-mass companions, left side) exhibit a broad range of higher velocities. A power-law fit (dashed line) suggests $v_{\mathrm{pec}}$ scaling inversely with the square root of total mass. This mass–velocity anti-correlation is evidence that more massive compact objects may receive weaker natal kicks on average, although observational biases and sample completeness must be carefully accounted for.}
\label{fig:kicks}
\end{figure}

This is codified in a statistically significant anti-correlation between the peculiar velocity on the one hand, and the binary's total mass $M_{\mathrm{tot}}$ or its orbital period $P_{\mathrm{orb}}$, on the other. In particular, the best-fit \rev{mass-}scaling is $v_{\mathrm{pec}} \propto M_{\mathrm{tot}}^{-0.5}$, shown in Fig.\,\ref{fig:kicks}. The orbital period anti-correlation (where short-period systems tend to have higher velocities, with $v_{\mathrm{pec}} \propto P_{\mathrm{orb}}^{-0.2}$) could reflect the effects of kicks in disrupting or tightening binaries. We note that these trends, while suggestive, must be interpreted with care due to observational scatter and selection biases (e.g. certain high-velocity systems might be harder to detect or measure if impacted by Galactic extinction). Nonetheless, these provide a benchmark for modelling natal kicks across the binary population.

Beyond the kinematics of known X-ray binaries, \textit{Gaia} DR3 has also enabled entirely new discoveries of compact objects via astrometric signatures. One of the exciting highlights is the detection of candidate \emph{dormant} \rev{(non-accreting, non-interacting)} BH companions in binary systems through astrometry alone. \rev{El-Badry et al. \cite{el-badry23}} and Chakrabarti et al. \cite{chakrabarti23}, for instance, identified a Sun-like star whose \textit{Gaia}-measured astrometric wobble and spectroscopic orbit indicate an unseen companion of $\sim 10~M_{\odot}$ – almost certainly a BH, now dubbed `Gaia BH1.' Alongside a handful of other candidates --- e.g., the even more massive `Gaia BH3' with $M_{\rm BH}$\,=\,33\,M$_\odot$ \cite{gaiabh3} --- an emerging population of stellar-mass BHs in wide orbits around normal stars are being discovered, found not via X-ray emission but via the gravitational tug detectable in astrometric data. Their discovery opens a new observational window on BH demographics, and \rev{extrapolations from these \textit{Gaia} findings suggest that our Galaxy likely harbours many hundreds of such non-interacting BH binaries \cite{chawla22}}. Going forward, refined astrometry, spectroscopic follow-up of promising candidates \cite{gandhi22, nagarajan25}, and additional data releases from \textit{Gaia} (and eventually its successors) will undoubtedly expand this inventory, providing critical constraints on binary evolution and the end stages of massive stars.

In summary, the \textit{Gaia} era has delivered transformative findings for compact-object kinematics. We now have direct evidence on a population level that BHs and NSs can both experience substantial kicks, and that high-precision astrometry can unveil even quiescent BHs. These results are a testament to the power of sub-mas--level astrometry in enriching our understanding of compact objects. At the same time, careful modelling and follow-up are required to account for biases in the astrometric data and to pin down the underlying physical mechanisms (e.g. the role of supernova vs. fall-back in BH kicks). We turn next to the other end of the mass scale of black holes found in galactic nuclei.

\section{Supermassive Black Hole Recoils}
While natal kicks of stellar remnants have been studied for decades, an analogous recoil phenomenon is predicted for supermassive black holes. When two galaxies merge, their central BHs are expected to form a binary and eventually coalesce. In the final moments of a BH--BH merger, if the gravitational wave emission is asymmetric, the remnant SMBH can receive a recoil kick. Numerical relativity simulations have shown that such gravitational wave kicks can be enormous: for certain configurations of BH masses and spins, recoil velocities of several 10$^3$~km~s$^{-1}$ are possible \cite{campanelli07}. Even more typical merger scenarios yield kicks of a few hundred km~s$^{-1}$. These speeds are comparable to or greater than the escape velocities of many galaxies, implying that a merged SMBH could be displaced from the galactic nucleus \rev{and eventually fall back}, or even be ejected into intergalactic space if the kick were sufficiently large.

Observationally confirming recoiling SMBHs is challenging. \rev{A completely isolated SMBH wandering through the intergalactic medium (IGM) is, in principle, detectable only gravitationally as a point lens for background sources, but the corresponding optical depth/event rate is likely minuscule given the rarity of such systems, the fact that the characteristic Einstein radius $\theta_{\rm E}=\sqrt{4GM_{\rm BH} D_{\rm LS}/(c^2 D_{\rm L} D_{\rm S})}$ will be of order fractions of arcsec or smaller for nearby galaxies (with distances to the source and the lens being $D_{\rm S}$ and D$_{\rm L}$, respectively, and their mutual distance being $D_{\rm LS}$ \cite{paczynski86}; $G$ is the gravitational constant), and finally because the characteristic variability timescale $t_{\rm E}\propto\sqrt{M_{\rm BH}}$ far exceeds typical survey baselines for SMBH masses, at least historically. Electromagnetic detection via Bondi(-Hoyle) accretion from the diffuse IGM is also unfavourable given the very low gas densities expected \citep{bondi52}.}

\rev{However, kicked black holes are expected to drag along material that is initially orbiting within an effective radius $r_{\rm k}$\,=\,$\frac{GM_{\rm BH}}{v_{\rm NK}^2}$, which is of order the size of the broad-line region in typical Seyferts \cite{merritt06}. This material may continue to be excited and light up the trajectory of the kicked source.} Several such intriguing candidates have been reported. A classic example is the quasar SDSS~J0927+2943, which exhibits double sets of broad emission lines in its optical spectrum. Komossa et al. \cite{komossa08} proposed that this could be explained by a recoiling SMBH: the idea is that the quasar's broad-line region remains bound to the SMBH after merger, so if the SMBH recoils at $\sim$\,2,650\,km\,s$^{-1}$, its broad lines would be Doppler-shifted relative to the host galaxy's rest frame. Alternative interpretations (such as a superposition of two merging active nuclei) exist, but clearly it is critical to test for kicked active galactic nuclei (AGN). More recently, \rev{Chiaberge et al.} \cite{chiaberge17} reported an observation of quasar 3C~186, which is located $\sim 8$~kpc from the centre of its host galaxy and whose optical spectral lines imply a velocity offset of $\sim$\,2,100\,km\,s$^{-1}$. The authors argue this is a strong candidate for an ejected SMBH, the result of a gravitational-wave recoil event in a merger of two massive BHs. \rev{Yet another case is that of RBH-1, a candidate supermassive black hole with a runaway velocity of $\approx$\,950\,km\,s$^{-1}$ found at the bow-shocked tip of a linear gaseous extension offset from its candidate host galaxy by 62\,kpc \citep{vandokkum26}. Such examples continue to be closely scrutinised and questioned \cite{sanchez26}, highlighting the degeneracies and community interest in this research.}

High angular-resolution imaging is key to confirming spatial displacements, whereas spectroscopy can reveal velocity signatures. Astrometry in the traditional sense (measuring proper motions) is generally not sensitive enough for individual distant SMBHs --- a BH kicked at 1,000\,km\,s$^{-1}$ would have a proper motion of only $\sim$\,0.2\,$\mu$as/yr at 1~Gpc, far below current capabilities, though this shift would cumulatively grow to become a detectable spatial offset with time. However, \textit{Gaia} is also sampling QSOs as stationary reference points across the whole sky; any significant proper motion of a quasar could indicate a nearby perturbation or recoil, although at present \textit{Gaia} limits for quasar proper motions are serving more as a check on systematics and emission from relativistic jets than a true measurement of quasar motion \cite{kovalev20}. The hope is that, in future, if a SMBH were ejected in a relatively nearby galaxy, high-precision astrometry might measure the reflex motion of stars or gas in the galaxy responding to the absent central mass, or detect an off-nuclear stellar cluster moving with the BH.

\section{Occultation Astrometry: A Fresh Perspective for High Angular-Resolution X-ray Studies}
Long before the era of \textit{Gaia} and VLBI, astronomers used occultation techniques to improve angular resolution. Such studies, dating back to the early days of radio and X-ray astronomy, utilised the orbit of the Moon as it passed in front of a distant source to record the time of occultation. Since the speed and exact location of the Moon are known to great precision, one can deduce the position of the source with far greater precision than the native resolution of the detector. 

In the early days of X-ray astronomy, localization of sources was a major challenge --- X-ray detectors on sounding rockets or early satellites had typical positional uncertainties of half a degree or worse. Lunar occultations provided a breakthrough. For example, the bright X-ray source GX\,3+1, discovered in the late 1960s, had its position refined to arcsecond accuracy through a series of lunar occultation observations in 1971 \cite{janes72}. This was a remarkable feat of astrometry for the era: achieving $\sim 1''$ precision in X-ray source position at a time when no X-ray telescope could directly image with anywhere near that resolution. Occultation astrometry was applied to several other early X-ray sources and even in radio astronomy to determine precise positions of radio sources. In essence, the Moon served as a natural astrometric calibrator.

While lunar occultation measurements gradually fell out of use once X-ray and radio telescopes achieved high intrinsic resolution (e.g. the \textit{Chandra} X-ray observatory provides sub-arcsecond imaging), the legacy of this method is noteworthy. \rev{It demonstrated that extraordinary centroiding precision could be attained by using time-domain proxies instead of direct spatial resolution. This was a clever way of optimising geometry and timing to extract maximal positional information.}

We may now be able to revive the concept for the modern era. \rev{The Artemis Programme, being led by USA's National Aeronautics and Space Administration,} aims to return humanity to the Moon within this decade (by \rev{2028}, according to latest projections). Concurrently, the \rev{European Space Agency} and several Asian nations are also advancing serious plans and precursor landing missions. The race back to the Moon is clearly on. Against this backdrop, there are novel proposals to use the Moon as a science platform, including for astrometry  --- this time not by waiting for serendipitous occultations, but by placing instruments on the lunar surface \cite{gandhi24}. 

A stable lunar-based telescope or interferometer could exploit the Moon's lack of atmosphere and slow rotation to achieve extremely stable long-term astrometric measurements in X-rays. Natural topography on the Moon (mountain tops or crater rims) could serve as occultors in-situ, with every object in the sky undergoing an occultation when it rises, and also when it sets, every lunar day. Further control should be possible through custom-made occultation devices (collimators, plates) which can be judiciously placed in the field-of-view to enable targeted measurements of specific objects of interest. 

The big advantage of (lunar) `ground-based X-ray astronomy' is that it mitigates several engineering challenges. No longer will it be necessary to build and launch telescopes all in one go; instead, it will be possible to accrue collecting area in a modular fashion and to implement repairs and improvements as necessary, with direct on-site access to the facilities. For detected source count-rates of several hundred per second, Gandhi \cite{gandhi24} shows that an occultation X-ray telescope can attain angular resolution surpassing that provided by the imaging \rev{finesse} of {\em Chandra}. Some challenges that will need to be contended with include lunar dust, thermal variations and seismicity; but these challenges will be faced, and dealt with, by all astronomical facilities expected to be sited on the Moon, and are not insurmountable. If feasible, this proposal opens up new horizons for high angular-resolution X-ray astrometry using well-understood timing techniques and detector technologies.

\section{Future Outlook}
High-precision astrometry is still in its infancy in many respects, with multiple facilities set to extend our reach, and with it our understanding of compact objects. On the observational front, \textit{Gaia} has an  upcoming Data Release~4 (expected in late \rev{2026}) and then a final Data Release~5 (at the end of the mission), which will further improve parallax and proper motion precision thanks to longer time baselines and updated calibrations, and will include more binary orbital solutions. Each data release will refine distances and velocities for NSs and BH binaries, allowing better mapping of their trajectories and inferences of their origins. We can expect a larger and more robust catalogue of astrometric BH binaries from \textit{Gaia} ---- confirming whether the few candidates found so far are the tip of an iceberg of dormant BHs in wide binaries. 

Looking further ahead, the {\em Nancy Grace Roman Telescope} will soon enable new studies of similar systems in the near-infrared through optically-thick gas and dust clouds \cite{gandhi23_roman}. \rev{Furthermore, {\em Roman}'s observational cadencing will be sensitive to microlensing events caused by free-floating stellar-mass BHs over wide areas of the sky --- a unique probe of isolated BHs not in binaries \cite{sajadian23}.} There are also proposals for \textit{Gaia} successors that would broaden capabilities including {\em Gaia-NIR} \cite{gaianir}, {\em Theia} \cite{theia}, and the approved Japanese mission {\em JASMINE} \cite{jasmine}.

Interferometry is also being advanced at multiple wavelengths. The VLTI/ GRAVITY instrument has already demonstrated $\sim$\,10\,$\mu$as astrometric precision in the near-infrared by observing the orbits of stars around the Milky Way's central SMBH \cite{gravity18}. Future interferometers, whether ground-based (e.g. the proposed Next Generation Very Large Array, ngVLA \cite{ngvla}, in the radio) or space-based in X-ray \cite{uttley21} will extend the reach in angular resolution enormously, but some of these technologies still await demonstration. \rev{The Square Kilometer Array (SKA) will be the next step in boosting radio sensitivities \cite{ska}. In terms of astrometric precision, SKA will advance the current frontier primarily through SKA-VLBI with phased SKA beams incorporated into global VLBI networks, ultimately reaching $\mu$as relative astrometric precision levels \citep{ska_vlbi}. This should extend microarcsecond-class astrometry to much fainter pulsars, X-ray binaries and transients than are accessible at present.}

In the realm of gravitational waves and multi-messenger astronomy, astrometry will play a supporting but important role, complementary to high angular-resolution imaging. \rev{As pulsar timing arrays open up the era of nanohertz gravitational waves likely from distant SMBH binary inspirals \cite{nanograv}, it is conceivable that astrometric monitoring of galaxies and quasars could pinpoint merging AGN in the nearby universe (e.g., by spotting a transient photometric centroid shift due to variability; \cite{chen22}). High-precision astrometry could also aid in studying suspected post-merger galaxies for signs of a recoil in their cores. Ultimately, such studies will enable a deeper understanding of supernova physics, compact object growth, and galaxy evolution.}

\begin{acknowledgement}
The author acknowledges Y. Zhao, C. Dashwood Brown, P. Charles, T. Maccarone, J. Paice, A. Rao, M. Johnson, P. Nuchvanichakul, C. Knigge, K. Belczynski and many others for stimulating discussions and collaboration on astrometric studies. The Science and Technology Facilities Council and the Royal Society are acknowledged for support. \rev{A special thanks to the organisers of the stimulating ISSAC-2024 symposium. The use of the Generative AI ChatGPT model 5 is acknowledged for syntax checks and linguistic improvements.}
\end{acknowledgement}

\bibliographystyle{spphys}
\bibliography{PGandhi}

\end{document}